\documentclass[acmtog, authorversion,nonacm]{acmart}
\usepackage[inline]{enumitem}

\usepackage{booktabs} % For formal tables
\usepackage{ccicons}  % For Creative Commons citation icons
\usepackage{todonotes}
\usepackage{subfig}
\graphicspath{{figures/}}
\usepackage[english]{babel}
\usepackage[autostyle=true,english=american]{csquotes}
\usepackage{gensymb}
\usepackage{balance}
\usepackage[most]{tcolorbox}

\usepackage{acronym}
\newacro{AI}[AI]{Artificial Intelligence}
\newacro{UI}[UI]{user interface}
\newacro{GUI}[GUI]{graphical user interface}
\newacro{TLX}[TLX]{NASA-Task Load Index}
\newacro{RTLX}[Raw-TLX]{NASA Raw-Task Load Index}
\newacro{ER}[ER]{error rate}
\newacro{TCT}[TCT]{task completion time}
\newacro{HCI}[HCI]{Human-Computer Interaction}
\newacro{UX}[UX]{user experience}
\newacro{HFE}[HFE]{Human Factors and Ergonomics}
\newacro{cuDNN}[cuDNN]{CUDA Deep Neural Network library}
\newacro{RMSE}[RMSE]{root mean squared error}
\newacro{HMD}[HMD]{Head-Mounted Display}
\newacro{RF}[RF]{Random Forest}
\newacro{GP}[GP]{Gaussian process, long-plural = Gaussian processes}
\newacro{KNN}[\textit{k}NN]{\textit{k}-nearest neighbor}
\newacro{NN}[NN]{Neural Network}
\newacro{DNN}[DNN]{ Deep Neural Network}
\newacro{CNN}[CNN]{Convolutional Neural Network}
\newacro{FCL}[FCL]{fully connected layer}
\newacro{BoD}[BoD]{Back-of-Device}
\newacro{FOV}[FoV]{field of view}
\newacro{RW}[RW]{Real World}
\newacro{IFRC}[IFRC]{index finger ray cast}
\newacro{FRC}[FRC]{forearm ray cast}
\newacro{EFRC}[EFRC]{eye-finger ray cast}
\newacro{HRC}[HRC]{Human-Robot Collaboration}
\newacro{HRI}[HRI]{Human-Robot Interaction}
\newacro{6DOF}[6DOF]{six-degree-of-freedom}
\newacro{LOOCV}[LOOCV]{leave-one-out cross-validation}
\newacro{CV}[CV]{cross-validation}
\newacro{RM}[RM]{repeated measure}
\newacro{ANOVA}[ANOVA]{analysis of variance}
\newacro{RMANOVA}[RM-ANOVA]{repeated measures analysis of variance}
\newacro{AGATe}[AGATe]{AGreement Analysis Toolkit}
\newacro{GHoST}[GHoST]{Gesture Heatmap Toolkit Gesture Heatmaps Toolkit}
\newacro{GREAT}[GREAT]{Gesture Relative Accuracy Toolkit}
\newacro{GRT}[GRT]{Gesture Recognition Toolkit}
\newacro{DTW}[DTW]{Dynamic Time Warping}
\newacro{LHRD}[LHRD]{large high resolution display}
\newacro{GEQ}[GEQ]{Game Experience Questionnaire}
\newacro{SPGQ}[SPGQ]{Social Presence Gaming Questionnaire}
\newacro{JND}[JND]{just-noticeable difference}
\newacro{SUS}[SUS]{system usability scale}
\newacro{CSCW}[CSCW]{computer-supported cooperative work}
\newacro{CAD}[CAD]{computer-aided design}
\newacro{MR}[MR]{Mixed Reality}
\newacro{CVE}[CVE]{Collaborative Virtual Environment}
\newacro{AR}[AR]{Augmented Reality}
\newacro{AV}[AV]{Augmented Virtuality}
\newacro{VR}[VR]{Virtual Reality}
\newacro{PRISMA}[PRISMA]{Preferred Reporting Items for Systematic Reviews}
\newacro{PRISMA-Scope}[PRISMA-ScR]{Meta-Analyses Extension for Scoping Reviews}
\newacro{TF-IDF}[TF-IDF]{Term Frequency-Inverse Document Frequency}
\newacro{TF}[TF]{Term Frequency}
\newacro{AVs}[AVs]{Automated Vehicles}
\newacro{eHMIs}[eHMIs]{external Human-machine interfaces}
\newacro{SAR}[SAR]{Spatial Augmented Reality}
\newacro{IFR}[IFR]{International Federation of Robotics}
\newacro{ADLs}[ADLs]{Activities of Daily Living}
\newacro{LED}[LED]{Light-Emitting Diode}
\newacro{DoF}[DoF]{Degree-of-Freedom} \newacroplural{DoF}[DoFs]{Degrees-of-Freedom}
\newacro{HHC}[HHC]{Human-Human Collaboration}
\newacro{IDF}[IDF]{Inverse Document Frequency}
\newacro{WHO}[WHO]{World Health Organization}
\newacro{IS}[IS]{Information System research}

\begin{document}

\title{Evaluating Assistive Technologies on a Trade Fair}
\subtitle{Methodological Overview and Lessons Learned}

\author{Annalies Baumeister}
\authornote{These authors -- in alphabetical order -- contributed equally to this research.}
\orcid{0009-0002-1007-7549}
\email{annalies.baumeister@fb4.fra-uas.de}
\affiliation{
    \institution{Frankfurt University of Applied Sciences}
    \city{Frankfurt}
    \country{Germany}
}

\author{Felix Goldau}
\authornotemark[1]
\orcid{0000-0003-4552-6842}
\email{felix.goldau@dfki.de}
\affiliation{
    \institution{German Research Center for Artificial Intelligence (DFKI)}
    \city{Bremen}
    \country{Germany}
}

\author{Max Pascher}
\authornotemark[1]
\orcid{0000-0002-6847-0696}
\email{max.pascher@udo.edu}
\affiliation{
    \institution{TU Dortmund University}%
    \city{Dortmund}%
    \country{Germany}%
}%
\affiliation{%
    \institution{University of Duisburg-Essen}%
    \city{Essen}%
    \country{Germany}%
}

\author{Jens Gerken}
\orcid{0000-0002-0634-3931}
\email{jens.gerken@udo.edu}
\affiliation{
    \institution{TU Dortmund University}
    \city{Dortmund}
    \country{Germany}
}

\author{Udo Frese}
\orcid{0000-0001-8325-6324}
\email{udo.frese@dfki.de}
\affiliation{
    \institution{German Research Center for Artificial Intelligence (DFKI)}
    \city{Bremen}
    \country{Germany}
}
\author{Patrizia Tolle}
%\orcid{}
\email{tolle@fb4.fra-uas.de}
\affiliation{
    \institution{Frankfurt University of Applied Sciences}
    \city{Frankfurt}
    \country{Germany}
}

\renewcommand{\shortauthors}{A. Baumeister, F. Goldau, M. Pascher et al.}

\begin{abstract}
%Why is the problem important?
%What have you done?
%What is new about it?
%What did you find?
%What does it implicate in the bigger picture?

User-centered evaluations are a core requirement in the development of new user related technologies. However, it is often difficult to recruit sufficient participants, especially if the target population is small, particularly busy, or in some way restricted in their mobility.
We bypassed these problems by conducting studies on trade fairs that were specifically designed for our target population (potentially care-receiving individuals in wheelchairs) and therefore provided our users with external incentive to attend our study.
This paper presents our gathered experiences, including methodological specifications and lessons learned, and is aimed to guide other researchers with conducting similar studies.
In addition, we also discuss chances generated by this unconventional study environment as well as its limitations.
\end{abstract}




\begin{teaserfigure}
  \includegraphics[width=\textwidth]{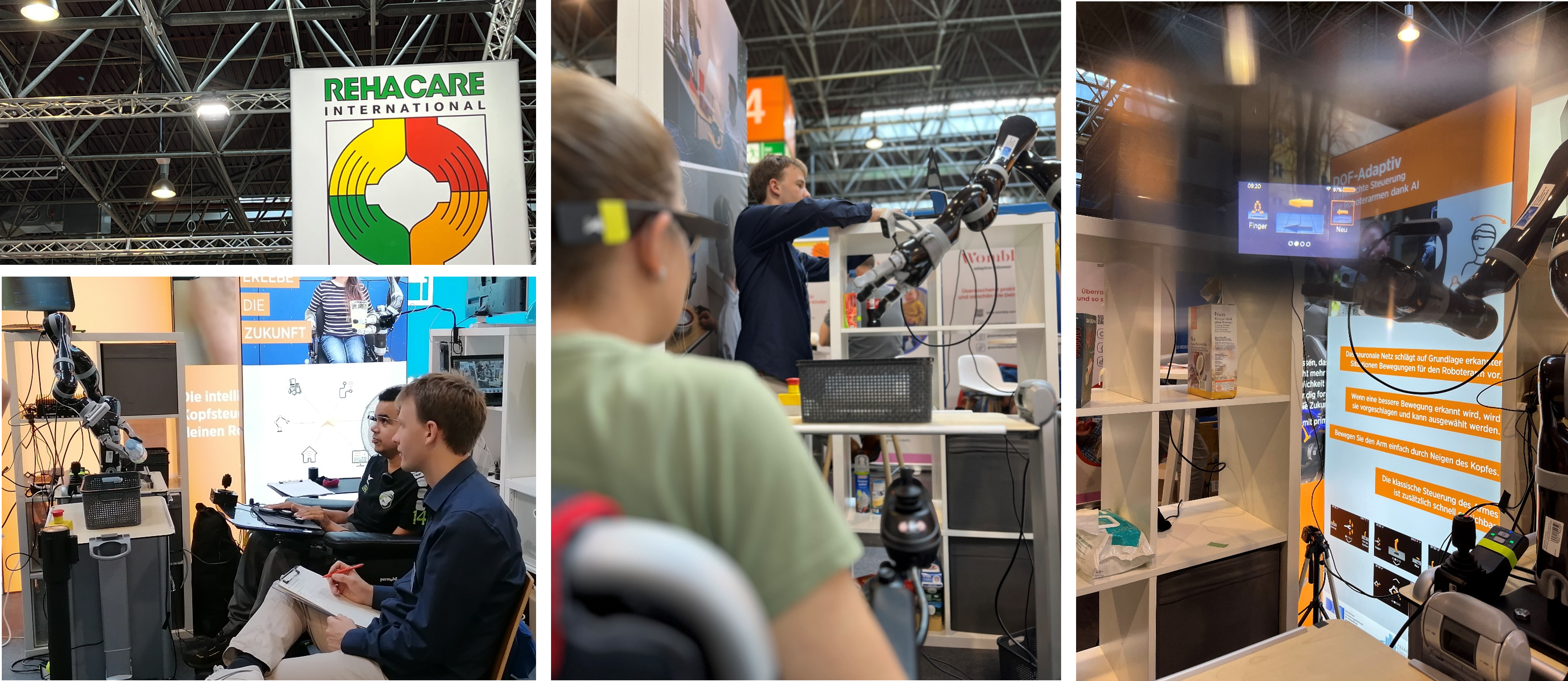}
  \caption{Impressions of a trade fair-based user evaluation}
  \Description{Four pictures in a collage, showing an expo booth with researchers and wheelchair-users controlling an assistive robot arm.}
  \label{fig:teaser}
\end{teaserfigure}

\maketitle

\section{Introduction: Why Run User Evaluations on a Trade Fair?}\label{sec:Introduction}

In 2023, the \ac{WHO} constituted 15\% of all people around the world to live with disabilities~\cite{who.2023}. In Germany alone, 7.8 million people were identified as severely disabled by the end of 2021, with over half suffering from physical impairments, significantly impacting their mobility and leading to social and professional exclusion~\cite{SchwerbehindertenStatistik.2022}. 
For many of these, who require consistent individual care-giving, assistive technologies can become an important tool to increase independence~\cite{Martinsen.2008}. These  range from basic aids to advanced robotics, offering independence and reducing caregiver dependency; thereby improving well-being and allowing individuals with disabilities to participate more fully in life. Furthermore, the aging population and their preference for aging in place amplify the demand for novel solutions~\cite{McMurray.2017}.

The development of advanced assistive devices, such as robotic arms capable of performing daily tasks, presents a new frontier in support for those with motor impairments~\cite{Connors.2019,Muthu.2023}. However, challenges in the user's control and associated stress with autonomy need careful management.
In these care-centered environments, flexibility and user-friendly controls are essential, as the technical proficiency of users varies widely and the complexity of such systems often poses barriers to effective use, especially for those with disabilities~\cite{Stephanidis.2019}. 
Current research in shared control and \ac{AI} aims to improve the usability and accessibility, emphasizing the importance of an intuitive operation and tailored user interfaces.
This focus on enhancing human-robot interaction underscores the broader challenges in this field's research and development, including ethical and logistical hurdles, safety, recruitment, and the diverse needs of users.

%Amid the rise of \ac{AI}-driven automation in areas like autonomous driving, content curation, and financial decision-making, it's crucial to acknowledge the indispensable role of human involvement. The essence of \ac{AI}'s success hinges on enhancing and extending human capabilities, not replacing them. This perspective is at the heart of human-centered \ac{AI}, which aims to develop \ac{AI} systems that support and augment human abilities, ensuring that technology meets human needs while enhancing autonomy and keeping humans informed and in control.

%This approach is particularly relevant for those dependent on assistance for daily activities. 
Emphasizing user collaboration, research highlights the importance of involving users as active participants in the design process, leveraging their unique insights into individual needs and experiences~\cite{Merkel.2018, Woodcock2020, Frennert2014}. This collaborative and interdisciplinary approach, supported by organizations like the WHO, underscores the value of viewing users as partners in the development and application of assistive technologies. Such involvement not only enhances functionality and accessibility, but also supports mental well-being and autonomy. 

This is particularly relevant for those dependent on assistance for daily activities. 
The \enquote{Design for All} philosophy~\cite{Stephanidis.1998}, integrating human-centered design with accessibility, advocates for incorporating user insights in the design process from the outset. %, including those with disabilities.
Based on ethnographic studies and direct engagement with our target audience, our research builds on this foundation, identifying specific needs and challenges to inform the development of assistive technologies that address physical, social, and collaborative aspects for a more inclusive, empowering solution.

% Most technical innovations need to be thoroughly tested to make any reasonable claim about functionality, user acceptance, or their general viability as a usable device. 
While this can, for some applications, be executed isolated in a controlled lab environment, most applications require interactions with users at some point during development; Some fields (e.g.\ \ac{HCI}, \ac{HRI}) even have these at the very core of their research.
However, sampling sufficient study participants to make reasonable claims is often difficult and not always a trivial task. This holds especially true for the field of assistive robotics, where the target population is limited in size and its members are often tightly scheduled and potentially vulnerable; be it physically, mentally, socially, or simply by introducing them to research-generated technologies that might help them, the production timelines of which are however too long to have any immediate use. 

Nevertheless, as the field has high potential of improving the lives of people with technology, a lot of interesting and promising research is conducted and evaluated. Yet one can discuss the generalizability of various studies, as either only small shares of the study participants actually belong to the target population, or the total cohort of users is very small.

%%% ToDo: List a few
For example, \citeauthor{Herlant.2016modeswitch} analyzed assistive robot control and compared the classic manual mode switching approach to one that is automatic and  time-optimal. Their study shows interesting results, especially regarding the challenges associated with mode switches. However only their initial interviews were conducted with users from the target population ($N=3$), whereas the rest of the evaluation was performed with able-bodied subjects~\cite{Herlant.2016modeswitch}. 
Similarly, \citeauthor{argall_sharedAutonomy} present an approach of assistive control using a body-machine interface and shared control. They tested it with 6 users, only one of which was a potential receiver of the technology~\cite{argall_sharedAutonomy}.
Positive examples exist too, but often require specialized cooperation between partners: For example, \citeauthor{argall_multiUsers} proposed an assistive optimization framework with humans in-the-loop and conducted a pilot study with 17 subjects, 4 of which had spinal cord injuries. However, this group is located at the \emph{Shirley Ryan AbilityLab}\footnote{Shirley Ryan AbilityLab (formerly Rehabilitation Institute of Chicago) \url{https://www.sralab.org/} last visited \today} and therefore profits from an established partnership with a rehabilitation hospital.
%%% Done: list a few

\subsection{Contribution}\label{sec:Contribution}
In this work, we present an alternative approach, which utilizes the attracting effect of care-related trade fairs: By conducting studies in a booth of the fair, we were able to reach a considerably higher number of potential users when compared to a classical lab study and thus gain valuable insights for our research. 
As these studies required strategies specifically tailored to this environment, we aim to share our expertise with the community. In our case specifically, the studies focused on evaluating (shared) manual control of assistive robot arms for wheelchair users with limited upper limb mobility.

We therefore present our experiences regarding:
\begin{itemize}
    \item opportunities, chances and advantages of trade fair-based studies,
    \item the special and/or unconventional requirements and preparations necessary to conduct such a study, and
    \item the limitations that arose in this unconventional environment. 
\end{itemize}

\section{Related Work: Participatory Evaluations of Assistive Technologies}
\label{sec:related}

%\begin{itemize}
%    \item Wichtigkeit der Einbringung der NutzerInnen (vgl Interact-Paper)
%    \item Studie auf Messen / Konferenz!
%    \item Spezialform von In-the-wild-Study~\cite{researchInTheWild}
%    \item Langfristige Studies (zuhause installieren)
%    \item Eigene vorangegangene Studien:
%    \begin{itemize}
%        \item Vergleich zur 2. AdaMeKoR Studie?
%        \item Vergleich zu AdaptiX~\cite{Pascher.2024adaptix}
%        \item Vergleich zu MobiLe-Rundreise (ethnographische studie)~\cite{Pascher.2021recommendations}
%    \end{itemize}
%\end{itemize}

Assistive technologies are increasingly recognized to be valuable tools in domestic care settings as they offer individuals with physical impairments the opportunity to regain a measure of independence by supplementing or reducing the need for ongoing human assistance~\cite{Park.2020}. Despite these benefits, the adoption and use of these technologies face challenges, including cases of non-acceptance and abandonment. Research by  \citeauthor{klein.2020}~\cite{klein.2020} and \citeauthor{Merkel.2018}~\cite{Merkel.2018} underscores the necessity of aligning these devices more closely with the specific needs and preferences of their intended users. Heeding this sentiment, \citeauthor{Vines.2013}~\cite{Vines.2013} suggest involving potential users early in the development process to enhance device acceptance and utility.

In the social sciences, such participatory approaches of research are common ever since Kurt Lewin developed the \emph{action research method} and gained increased importance in health care research with the \ac{WHO} \enquote{Health for All Strategy}. Here, the assigned goal is an active participation of those affected in the research process, thereby collaboratively gaining knowledge, reflecting, influencing, and thus changing the research process~\cite{Baum854, vonUnger2007Aktionsforschung}. 

This synergy is also known in more technical fields, where they found the active participation of potential users in the design process of assistive technologies to be crucial but challenging~\cite{Dalko.2023,Montague.2015,Kulp.2018}: \citeauthor{Dalko.2023} highlight the significant difficulties in patient involvement, particularly among those with long-term illnesses, due to hierarchical barriers in care-institutions and a lack of established patient groups. Nonetheless, this involvement is key to developing devices that meet the specific needs of users and facilitate better outcomes in terms of usability and acceptance~\cite{Dalko.2023}. Towards the end of the 1990s, action research was introduced to \ac{IS} research, among others, by \citeauthor{Baskerville1996} and found its way into today's \ac{HCI} research. \citeauthor{Baskerville1996}~\cite{Baskerville1996} and \citeauthor{Hayes.2011}~\cite{Hayes.2011} both note that the participatory and collaborative approach of action research fits to methods and issues previously used by researchers in \ac{IS} and \ac{HCI} but extends their setup with an ethical framework.

In these technical fields, laboratory studies are very common and provide standardized and methodically controlled approaches, as well as being simpler and more economical by avoiding difficulties that could arise in the field.
However, data generated in laboratories misses everyday conditions and consequently leads to discussions of data validity~\cite{Baum854, DorschPsychoLexikon}. 
In addition, these in-lab studies often face logistical challenges, especially when involving participants with mobility impairments. The difficulties in transporting individuals to and from study locations can significantly impact the feasibility and cost of research, suggesting a need for more accessible and inclusive research methodologies~\cite{Schultz1992,Sonne2015,McMillan2015}.

Field research on the other hand, refers to processes that are observed in real life's  everyday settings, thus avoiding various issues inside laboratories. Downsides of field research however lie in the variation of conditions, the randomization of the perturbing conditions and therefore the method, as well as multiple (uncontrollable) effects that might  limit internal validation of such studies~\cite{lamnek2016}. Here, action research provides tools to methodically gain valid knowledge whilst collaborating with the target group~\cite{Hayes.2011}.

One variation could be conducting studies within the homes of participants, as this allows for a more realistic understanding of how assistive technologies function in everyday settings while keeping the environment semi-controlled. However, these studies typically involve smaller sample sizes due to logistical and financial constraints, potentially limiting the generalizability of findings~\cite{Demers2016,Anderson2013}.

Still, research conducted in real-world settings (in-the-wild studies) provides valuable insights into how assistive technologies are used in daily life. These studies can highlight issues of device acceptance and long-term use that may not be apparent in more controlled research settings~\cite{Montague.2015,Kulp.2018,Goulden.2017}.

In summary, the design and development of assistive technologies benefit significantly from involving the target user group at every stage, from ideation to final product testing. Identifying this requirement ensures that the resulting devices are not only technically sound but also tailored to the real-world needs and preferences of their users. Addressing the challenges associated with in-lab, in-home, and in-the-wild studies is essential for advancing our understanding of assistive technology use and improving outcomes for individuals with disabilities.

\section{Methodology of our Example User Evaluations}
\label{sec:method}
A very important element of any user-centric evaluation is subject recruitment, the success of which depends primarily on one's location: It can be very struggling to sample a sufficient number of people from the target population if the study is location bound and the subjects are expected to travel to the research lab. 
As an alternative, we searched for places with a considerably higher-than-usual distribution of care-receiving individuals, finally landing on trade fairs for care and rehabilitation. 
Participants from previous studies suggested to look at these, as they are fixed annual events within the community.

Here, healthcare providers and (self-advocated) societies gather to exchange experience and inform the public, among other topics, about provided services, relevant regulations, and available federal social-care benefits. 
In addition, they also include a large marketplace for manufacturers to showcase their (new) designs and technologies to the target audience. 
This creates large incentives for those involved in care, as they can personally observe and experience a large number of products which might have the potential to improve their lives.
The condensed experience is especially attractive for people who find traveling to be particularly strenuous, be it due to disabilities or other means.

Ultimately we decided on the \emph{REHAB} trade fair\footnote{\emph{REHAB} trade fair. \url{https://www.rehab-karlsruhe.com}, last retrieved \today} in Karlsruhe (Germany) and the \emph{REHACARE} trade fair\footnote{\emph{REHACARE} trade fair. \url{https://www.rehacare.de}, last retrieved \today.} in D{\"u}sseldorf (Germany), both internationally well known trade fairs for rehabilitation and care. 
Consequently, they are also known to be visited by many people with a disability who not only use it to inform themselves about new aids, but also to meet up and network with their peers. 

For us, this meant extremely high numbers of potential (primary and secondary) users, which do not need to explicitly travel only to join our studies, but were basically already on-location. For people who were previously associated with our projects (e.g.\ due to the participatory design), we were able to offer discounts on the entry.
These conditions allow for a way tighter definition of the study subjects, simply because of the associated shift in the local distribution: Instead of rough estimates of potentially care-receiving individuals, substantial sample sizes can be reached with tighter and more fitting inclusion criteria, e.g.\ acquiring only wheelchair users with limited mobility in their upper extremities. 

However, this special environment also greatly influences the objectives that can reasonably be evaluated: We have to assume our participants to be less focused, both due to  more external interference, as well as individual agendas as trade fair guests. The latter also strictly limits the available time per user.
As a result, it is reasonable to focus on qualitative objectives, relying more on interviews and personal user remarks, rather than interpreting too much into individual trials. 
%For study A, which was mainly intended as loose survey and a user-recruitment opportunity, this fit perfectly. However, as study B was designed as a final evaluation, we also wanted to generate at least some quantitative data. We therefore dismantled our scenario (retrieve multiple objects from a shelf) into smaller distinct tasks (approach, grasp, and pull back a single object) that were easier to compare and were less likely to be interrupted or excessively influenced by the trade fair environment.
For us, this meant selecting objectives that focus on user feedback, acceptance and preferences.

\subsection{Experimental Design}\label{sec:exp_design}
As the main reason to select this type of study is a shortage of suitable participants, the most valuable resource whilst conducting the study is the users' time and willingness to contribute.
We therefore aimed to optimize their experience as far as possible, by isolating temporal bottlenecks and widening them by conducting the study partially in parallel using multiple researchers: One person recruits the next user and give introductory information, while another one performs the robot interaction with the current user, and a third person debriefs the previous user and runs an final interview.

%each study involved multiple people. For the simpler study A, two people sufficed, each of which would guide an individual participant through interviews and the practical part; whilst being delayed to the other person to achieve a maximum time-wise utilisation of the study setup. Partially, a third person was used to attract new participants

%For the more ambitious study B, the local team was slightly increased: At least one person was consistently recruiting new people from the trade fair guests, two were conducting the technical part of the study (one of which was taking qualitative notes of user statements or situations), and two were interviewing the participants before and after the technical experiments.

\begin{figure}[hbt]
    \centering
    \includegraphics[width=\columnwidth]{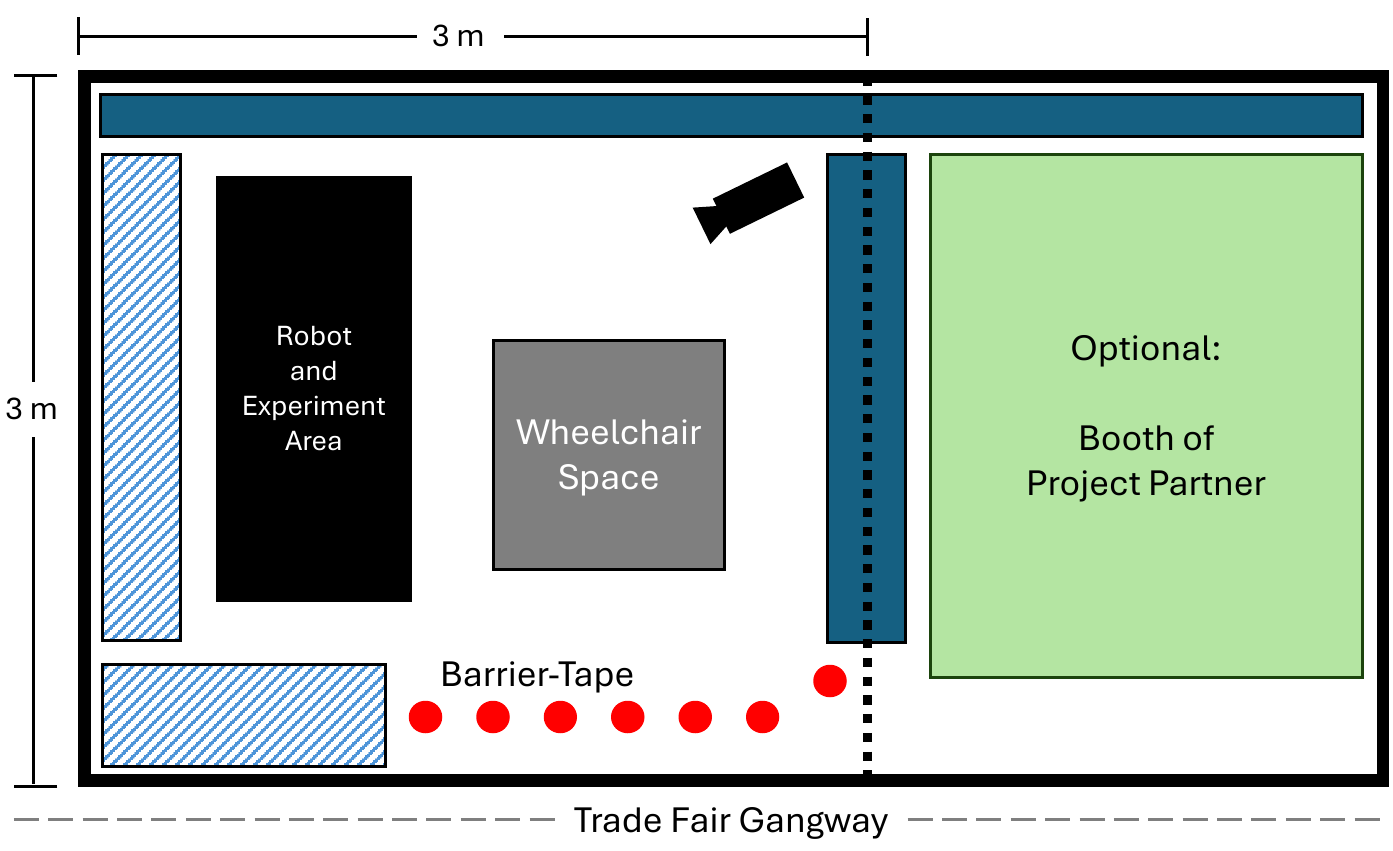}
    \caption{Sketch of exemplary booth setup}
    \Description{A rectangle showcasing a top-down view of a trade fair booth.}
    \label{fig:aufbau}
\end{figure}

The physical setup of our studies at trade fairs was relatively minimalistic: a robot arm is mounted to a table in a standard booth with sufficient room around it, such that both wheelchair users as well as researchers can easily access the system.  A simple sketch of an exemplary booth setup is shown in Figure\ \ref{fig:aufbau}: Our main experimentation area is shown on the left, with the right side representing our project partner's booth, which we could partially use for the interviews.
For the participants, we assured a minimal distance to the robot as a safety measure to the semi-chaotic nature of the environment. 
We purposefully avoided completely enclosing structures on the booth to preserve the visibility of the study as an advertising element for recruitment.
Finally, cameras and microphones were set up around the experimentation area to allow a subsequent analysis of user feedback and remarks in addition to notes taken by hand.

In addition to technical and logistical preparations, a sufficiently thorough ethical evaluations was necessary.
This included limiting the trial time to not cause exhaustion for our participants, which we attempted by structuring the trials such that the duration for every participant would not be longer than 60 minutes all together. 

Another topic of ethical review was privacy and data security during the collection of personal data in such a semi-public space. As previously mentioned, completely enclosing the booth for this did not seem a realistically implementable scenario, as it would result in higher costs (e.g.\ building materials, extra booth space) and a less attractive booth, thus making recruiting more difficult.
Instead, we used natural obstructions (e.g.\ semi-transparent shelves, striped blue in Figure\ \ref{fig:aufbau}) and planed to positioned the setup in such a way that the participants were oriented with their back towards the aisle and most other visitors passing by. 
Personal data was collected with a questionnaire to be filled out directly by the participants or their companions. 
%To make sure that participants could give informed consent during the first briefing, despite the expected distractions and high noise levels at a trade fair, all researchers involved were sensitized to the situation in advanced and were instructed to pay extra attention to it. 
Ethic approvals were obtained for both studies.

\section{Example Studies}\label{sec:studies}
We executed two very distinct studies on trade fairs at very different stages of projects: 
Study A was conducted at the \emph{REHAB} trade fair with a minimalistic physical setup and served mostly for an evaluation of the state of the art, harvesting of subjective user requirements, and recruitment of primary users of assistive robotics for the ongoing participatory research.
Close to the very end of our project, Study B was conducted at the \emph{REHACARE} trade fair in D{\"u}sseldorf (Germany) and served more as the final stage of evaluation: We used this chance and our experience from study A to assess our solution with subjects sampled solely from the target population.
%For this, the technical functionality was previously ensured in independent studies with able-bodied participants~\cite{heuristicControl, Kronhardt.2022adaptOrPerish, Pascher.2023inTimeAndSpace,Pascher.2024devices}. 
% A detailed analysis of the results is shown in~\citeAn{ROMAN2024}.

This section provides a brief overview of the studies, focusing on their comparative user requirements, meta results and resonance.  

\subsection{Study A: Explorative User Tests}\label{sec:StudyA}
Having had only very limited experience of studies during a trade fair, we cautiously designed this first study to be mainly explorative, focus on qualitative user feedback, and potentially recruit people to join us for our upcoming participatory development.
% ~\citeAn{SocialRoboticsJournal}.
In this initial study, we aimed to compare existing options of manual control for an assistive robot arm by gathering feedback from people associated with care. This included care recipients, as well as informal and professional caregivers. We intentionally set these relatively loose conditions, as we were looking for diverse perspectives on the matter and were uncertain of the actual user distribution at the trade fair.

In total, 26 participants joined our experiment, 10 of which were care-receiving wheelchair users.
As we conducted the study in a booth during the 3-day \emph{REHAB} trade fair and each subject stayed with us for 30 to 45 minutes, this sample size came close to our full capacity and exceeded our previous expectations.

The results showed a preference for a direct control mapping combined with a minimization of mode switches, as well as a willingness to be confronted with more complex input devices and train with them.
In addition, we gathered invaluable insights into requirements and situations of our target population; By basically sampling from the wild, we were able to include user profiles that might have otherwise be forgotten, overlooked, or incorrectly excluded for our studies (e.g.\ users with spastics).

All together, the participants were all very interested and reported enthusiasm in joining our study. 
They shared previous experiences with similar systems, as well as contextual anecdotes, with the professional care-givers often providing technical expert clarifications.

\subsection{Study B: Final User Evaluation}\label{sec:StudyB}
Based on the positive resonance and high number of participants of the previous study, we decided to also conduct the final evaluation of our project-developed shared control approach at a trade fair.
% ~\citeAn{ROMAN2024}. % This time, 
In this study, we selected the larger \emph{REHACARE} and defined the inclusion criteria to be more specific: wheelchair-users with limited mobility of their upper limbs. Impressions from the study can be seen in Figure\ \ref{fig:teaser}.

Compared to the previous study, this evaluation was more structured and less explorative in order to allow us a more substantiated analysis of our shared control. While the concept was  previously shown to be functional with able-bodied users~\cite{Pascher.2024devices, heuristicControl}, the verification with the target group was still lacking. Bridging this gap was the main goal of the study presented here: 
Can the users learn the control sufficiently quick; and do they perceive it well and accessible? 
By adjusting to people's needs, we also analyzed the generalizability of the control to different input devices.
%At the trade fair, we partially set up walls around a booth to enable semi-private task execution with less distractions. The users would get recruited and instructed in advance, guided through the experiments afterwards, and finally interviewed and debriefed separately by a team of researchers.
In contrast to Section\ \ref{sec:StudyA}, the tasks were defined with measurable brief goals such that trials could be recorded and (partially quantitatively) compared afterwards.

Due to the more complex nature of the study and including interviews and preparation, participants spend roughly 60 minutes with us. In total, we managed to gather data from 24 people of the target population (wheelchair users with limited mobility in their upper limbs) during the 4-day long trade fair. 
As the physical capabilities of users varied vastly and were not known by us in advance, spontaneous adjustments to the mechanical setup were often necessary. 

Nevertheless, participants were again very enthusiastic and even glad to be included in the research process. 
The results showed the capabilities of the proposed control, as well as an increase in understanding and acceptance of the control during the trials. 

\section{Lessons Learned}
Trade fairs are messy: There is a lot of noise, it is at times very crowded, and there are various distractions. For example, spotlights disrupt visual interfaces, while the huge number of transmitting technical devices interfere with wireless connections.  But with the challenges do come opportunities and unexpected results that would not occur in more controlled setting. 

\textbf{Running Trials with an Audience:}
During both studies, seeing the robot in action was very attractive for other guests, who stopped to watch, often asking questions or giving comments, however also partially generating unwanted performance pressure for study participants. 
Consequently in Study B some participants remarked on the audience or other distractions during the interviews, implying that a calmer setting could have led to a better performance. One participant stated \enquote{\textit{But at the trade fair, there are people, time pressure. (\dots) and that's a bit more strenuous than at home.}}\footnote{This and all further direct quotes are translated from German.}  Another one said: \enquote{\textit{Oh, if you try that a few times and there's no audience there, (\dots) you'll become more confident.}} Other participants did not mind being watched during testing, with some even asking their companions take pictures or record them. 

\textbf{Acquisition of Participants:}
At both trade fairs, experience showed that guests are keen to test new technologies and are generally open to new ideas. We especially found our  target audience of wheelchair users with limited mobility in their upper limbs to be very curious and open to us, which greatly simplified and accelerated acquisition. Many participants expressed their joy in testing our robot control: \enquote{\textit{I just think it's really great that this option exists. And it was nice to be able to test it out.}}
Confirming our assumptions, participants told us that visiting these events is a regular (mostly annual) activity for them: They use it to stay on track with technology, find new assistive devices and gather with their peers.

Recruitment for the study was therefore relatively unstructured: by basically  sampling in-the-wild, we approached potential participants and had short condense introductory talks to get them interested in the study. As both the participants and our time on the trade fair was limited, the pre-study briefing was held minimalistic, especially when compared to recruitment talks in a lab-based study with travel time. 
While this interchange was at all possible due to the reduction of hurdles and consequently quick launches into participation, unexpected difficulties also arose due to diverse and previously unknown user situations. This included spastics which made holding controllers difficult, head rests or vision impairments that prevented the use of smart glasses, breathing aids or speech impediments that restricted communication, and neurological impairments. For most users, we were able to find spontaneous workarounds (e.g.\ repositioning and propping up controllers to lessen spastics, or setting up an external screen for vision impairments), but some trials had be aborted. In addition to blocking valuable time, this often left participants and researchers unsatisfied. %On the other hand did we gain new insights, for example about additional requirements for the smart glasses, because of those unexpected challenges.  

Specific to the trade fairs, we observed that appointments for trials tended not to work. The agenda of the guests changed too rapidly, as to allow them to return to our booth at a predefined time. In addition, it has to be considered that people participating in the afternoon after several hours at the trade fair tended to be more exhausted.

%\begin{itemize}
    %\item Bedingungen vor Ort: Messen sind laut, voll, viel Ablenkung Scheinwerfer / Deckenlichter störten visuelle Darstellungen. Technische Störungen (Viele funkende Systeme, WLAN). Zuschauer (inkl. Fotos, tlw. ungewollter Performence-Druck auf die User).
    %\item Aber auch: Viele aus der Zielgruppe da, viel Neugier und Offenheit, förderte Akquise
    %\item unexpected difficulties due to previously unknown and more diverse user situations (breathing aid, sprachstörungen, kopfstütze behindert aufsetzen von brillen, sehbeeinträchtigungen, ggbnfalls neurologische beeinträchtigungen...) $\rightarrow{}$ unstrukturiertere (weniger gezielte) Rekrutierung durch das sample from the wild, daher auch ein geringeres Vorgespräch wo vieles eindeutiger geklärt werden konnte. 
    %\item Leichter Zwang (der Researcher) auch alle/jede:n auf der Messe zu nehmen, die sich zur Teilnahme bereit erklären?
    %\item Distraction: People sometimes look around (can also happen to every person)
    %\item But it also has to be considered that people participating in the afternoon after several hours at the trait fair tend to be more exhausted.
    %\item Appointments did not work - personell fair agenda does not fit / does change too quickly so you cannot ensure them to be there on time    
%\end{itemize}

\textbf{Interviewing on a Trade Fair:}
The interviews of study B were conducted without an extra booth or fixed place. They mainly took place in a corner of our partner's booth and in some cases even at the edge of the aisle next to our booth. Not interviewing in our booth had the advantage that the next person could start their trial while the previous one was still in  their closing interview. 
For the researcher, it was inconvenient to constantly search for a place to conduct the interview, however, it provided the participants with a moment to clear their mind before the questions started. Further challenges that arose due to the conditions of the fair included the noise level, crowded space, as well as trials or demonstrations of our project partners in clear view. 
This led to a lot of distractions and interruptions during the interviews and made it hard for both parties to stay focused. Sometimes, caused by the background noises, the interviewers missed parts of an answer, thus loosing the opportunity for follow-up questions. This noise level also affected the transcription of the audio recording, resulting in 5 interviews, where part of the answers could not be transcribed.
In 6 further interviews, it was challenging to distinguish the speaker.  Although an \ac{AI}-based software was used for transcription, the transcripts needed to be manually corrected more thoroughly than usual. To find a practical solution to conduct interviews with less noise on a trade fair is not that easy. Still, a bigger booth with a fixed place for the interviews might have led to fewer distractions and a better interviewing quality.
\section{Discussion}
Typically, similar studies are conducted in laboratories of research institutes, i.e.\ subjects are recruited and invited in advance and the environment is known and completely controllable by the researchers. This comes with various advantages, accumulating to generally more predictable procedures: 
\begin{enumerate*}
    \item There are none to very few external influences in a lab, resulting in less distractions and consequently cleaner trials, 
    \item the participants are known and associated possible complications can be surmised in advance, and
    \item it is possible to schedule participants, as they will arrive independently and no study-related equipment needs to be transported.
\end{enumerate*}

However, this requires a pre-existing cohort of subjects to sample from (which is non-trivial for sufficiently tight requirements) and loads the burden of traveling to the lab onto the participants. Especially for people with physical impairments, this can be a major task, which, among other things, involves the availability of accompanying persons, options of transport, and space in one's own timetable, which is often enough stuffed with therapies.

In contrast, the presented trade fair-based approach inverts the situation. Both the environment as well as the participants are unknown variables; however, the latter have already traveled to the study location, such that they are not additionally burdened. This requires, of course, an existing trade fair that specifically addresses the target group. 

Other imaginable alternatives are purely virtual off-site studies (e.g.~\cite{Pascher.2024adaptix}), purely ethnographic studies where researchers travel to users' homes without equipment (e.g.~\cite{Pascher.2021recommendations}), or expensive and complicated evaluations where researchers visit users' homes and bring along equipment such as robots (e.g.~\cite{SocialRoboticsJournal}). However, each of these variations come with extensive downsides. 

Consistent throughout these alternative methodologies is the requirement of known predetermined users. In contrast, the presented studies recruited participants on-the-fly from a cohort of \enquote{free-roaming} trade fair guests.
This resulted in a much larger diversity of participants, both in terms of physical capabilities, as well as previous technical experience and acceptance, thus improving the scientific significance.
In particular, this includes less tech-savvy users who might be more critical towards such systems and would therefore regularly not get involved with technical research~\cite{tech_savvy_1, tech_savvy_2}. 

On the downside, this came with unforeseen challenges (e.g.\ partial blindness on one eye) which required spontaneous adjustments, not all of which could be met on-site. Still, providing a larger group of potential future users with a chance to evaluate the technology meant that, in the spirit of action research, as many people from our target group were involved in our final evaluation as possible: They shared their knowledge as experts in their own field, including former experiences with assistive technologies, life situation and the resulting requirements; thus supplementing their perspective and influencing further research~\cite{Wright2013, lamnek2016}. %Furthermore, to do research where the people are, made it possible to secure an individual and collective participation  of our target group in a crucial phase of our research~\cite{lamnek2016}. 

In Section~\ref{sec:exp_design}, we discussed the situation regarding privacy and data security of personal data and its impact on the design of the booth. Arriving at the trade fair (for study B) and seeing the situation on site for the first time, it soon became clear that the set up could not be implemented as planned. Instead of placing the demonstrator in a way that participants would face the back wall of the booth, they would now be positioned lateral and be more visible for other passing visitors. That meant less privacy and more distractions during the trial. 
However, it is quite common on such trade fairs for visitors to try out new aids (e.g.\ wheelchairs or robotic eating devices) in public and without any consideration to privacy. 
Doing so, visitors of the trade fair are well aware that others stop to watch or even take pictures. Still, whenever we noticed someone taking pictures, we asked them to only depict the robot and ensure the participants to not be recognizable, e.g.\ by taking the picture from behind. 
We therefore conclude that the reduction of privacy at a trade fair is a reasonable circumstance for the participants. 

Another challenge was a secure way of collecting of personal data. Even though some participants were able to manually fill out the questionnaire (sometimes with assistance from their companions), they often requested a researcher to collaborate.  This was done as discreetly as the circumstances allowed.
%But because of the noise level and action around us, passing visitors were to distracted to catch anything anyway. Than again we did not asked for contact information, like home address or phone number. Participants could leave their e-mail if they liked to be further contacted by us. This than was as discreetly collected as possible, too. 
The trials itself took longer than anticipated: Instead of planned 45 minutes, most participants stayed with us for about 60 minutes. Luckily, this showed not to be a problem at all, as all participants were happily ready to invest this time. Some expressed excitement and joy during or after their trail.  %Therefore, after the first few trails, we rather planed with an our for each participants than rushing them through the study.

\textbf{Study Result Validity:}
In our investigation of inclusive and assistive human-robot interaction, we conducted two distinct studies, each offering unique insights into the effectiveness and acceptance of our approach.

The first study, Study A, took place at the REHAB trade fair in Karlsruhe, Germany. With a focus on qualitative feedback and exploration, we engaged 26 participants over the course of three days. 
Building upon the success of Study A, we proceeded to Study B, a final evaluation conducted at the REHACARE trade fair in Düsseldorf. Here, we targeted wheelchair users with limited upper limb mobility, totaling 24 participants over four days. Adopting a more structured methodology, including interviews and defined task trials, we sought to validate the effectiveness and acceptance of our shared control approach. 

Together, these studies provide a robust foundation -- involving 50 participants -- for assessing the viability of assistive robotics in inclusive environments. With a diverse participant pool and a combination of qualitative and structured methodologies, we have garnered valuable insights into user preferences, requirements, and acceptance, paving the way for future advancements in inclusive human-robot interaction.
In general, participants demonstrated enthusiasm and willingness to engage with the research process, reinforcing the validity of our findings.

\section{Conclusion}
We presented an in-depth analysis of the capabilities that arise from running robotic studies at trade fairs, with a special focus on assistive technologies designed for care-receiving individuals.
For this, we showed a generalized methodology, provided brief summaries of the approaches and results of two different studies we performed on trade fairs, and discussed our experiences with this unconventional study setting.

As discussed, the study conditions on a trade fair differ vastly from those of a typical lab-based evaluation. In short, one exchanges a bit of predictability and general control in the lab with a way better adjusted localized target population (i.e.\ easier recruitment of appropriate subjects) and a more realistic in-the-wild environment. As shown, the setting of a trade fair has its own, partially chaotic, dynamics. Therefore a thorough planing is needed, including situation-aware preparations, but also researchers that are willing to react spontaneously and be ready to improvise. 

This definitely does not expose trade fairs as the general go-to option for user evaluations, but instead highlights them as a valid alternative to a lab-based study. This is especially the case for studies that are mostly qualitative and focus on a user group that is either limited in size, has difficulties traveling to research laboratories, or difficult to contact in the first place.

While this requires a mobile study setup, custom preparations in advance, and capabilities for spontaneous problem solving at the trade fair itself, we can highly recommend examining this option. For us, it provided larger numbers of more appropriate participants and extremely valuable insights from exchanges with the real target group. Besides, quite a lot of our participants stated, that they had fun during the trails and enjoyed being able to join the study and partake in our research. 

%\begin{itemize}
%    \item Würden wir das wieder machen?
%    \item Vorteile / Nachteile / Nächstes mal anders?
%    \item extremely valuable insights from the real group
%\end{itemize}

\begin{acks}
This research is supported by the \textit{German Federal Ministry of Education and Research} (BMBF, FKZ: \href{https://foerderportal.bund.de/foekat/jsp/SucheAction.do?actionMode=view&fkz=16SV8563}{16SV8563}, \href{https://foerderportal.bund.de/foekat/jsp/SucheAction.do?actionMode=view&fkz=16SV8564}{16SV8564}, and \href{https://foerderportal.bund.de/foekat/jsp/SucheAction.do?actionMode=view&fkz=16SV8565}{16SV8565}). Our studies were approved by the Ethics Committee of the \textit{Faculty of Business Administration and Economics of the University of Duisburg-Essen} and the Ethics Committee of the Carl von Ossietzky University of Oldenburg.
\end{acks}

\bibliographystyle{ACM-Reference-Format}
\bibliography{MainPaper}

\end{document}